# Long-term stability-degradation analysis of DUNE SiPMs in Liquid Nitrogen

**Thomas Tsang,**[a,1] **Flavio Cavanna,**[c] **Hucheng Chen,**[b] **Shanshan Gao,**[b] **Lingyun Ke,**[b] **and Koloina Rambeloson**[d]

[a] *Instrumentation Department, Brookhaven National Laboratory, Upton, NY 11973, U.S.A.*
[b] *Physics Department, Brookhaven National Laboratory, Upton, NY 11973, U.S.A.*
[c] *Physics Department, Femi National Laboratory, Batavia, IL 60510, U.S.A.*
[d] *Department of Physics, University of Antananarivo, Madagascar*

*E-mail*: *Tsang@bnl.gov*

ABSTRACT: We developed a cryogenic photon readout system to monitor arrays of DUNE SiPMs operating at various over-voltages in liquid nitrogen ($LN_2$) for over three months. Photoelectron signals were read out simultaneously via weak capacitive coupling to the µFEMB, a 32-channel charge sensitive readout board designed for 77 K to 300 K operation in liquid argon time projection chambers (LAr-TPCs). A waveform snippet acquisition scheme was implemented to capture waveform signals whenever they exceeded a predetermined trigger level; consequently, empty waveforms were discarded. Selected SiPM parameters were monitored to detect any deviations beyond statistical fluctuations. While some parameters exhibited ~1σ variations over the 3-month test period, there was no evidence of drift being enhanced when operating at 4 V or 5 V compared to a 3 V over-voltage. However, a discernible PDE drop was observed across all channels, which warrants further investigation. In addition, we demonstrated that our system can simultaneously read out single photons from a selected group of SiPMs at room temperature.

[1] Corresponding Author.

## Contents

## 1. Introduction

The Deep Underground Neutrino Experiment (DUNE) is designed to conduct a broad exploration of neutrino physics over a wide range of energies with unprecedented detail. Liquid argon time-projection chamber (LArTPC) technology has been selected for the DUNE Far Detector (FD) experiment. To take advantage of the excellent scintillation properties of liquid argon (LAr), the FD will be instrumented with photon detectors (PDs) to provide precise timing and additional calorimetric information. The DUNE FD will consist of two modules, FD1-HD and FD2-VD, housed in independent cryostats each containing approximately 17.5 kt of LAr. Two additional modules (FD3 and FD4) are planned for a second phase of the project. Both the FD1-HD and FD2-VD designs utilize X-ARAPUCA PD technology; these optical modules act as light traps, capturing 127 nm LAr scintillation light and shifting it to ~450 nm. The resulting photons are trapped inside boxes with highly reflective internal surfaces until they are detected by silicon photomultipliers (SiPMs) distributed evenly around the perimeter of the PD module.

### 1.1 DUNE SiPM

In FD2-VD, the PD system is composed of 672 large-format X-ARAPUCA optical modules, each featuring 160 SiPMs. In contrast, FD1-HD implements a larger number of smaller modules (1,500), each containing 192 SiPMs. This results in a total of 107,520 and 288,000 SiPMs for FD2-VD and FD1-HD, respectively.

An extensive SiPM qualification program at cryogenic temperatures was conducted by the DUNE Photon Detection System Consortium [1]. The final models selected for FD1-HD were the FBK NUV-HD-Cryo 3T (6×6 mm², 50 µm pixel) and the HPK S13360 series (6×6 mm², 75 µm pixel). Both models demonstrated stable response times and similar performance in gain and correlated noise at slightly different over-voltages (OV). The same models were also selected for FD2-VD. For the 320 X-ARAPUCA modules on the LArTPC cathode, which exclusively use HPK S13360 SiPMs, a higher OV is currently under consideration to increase both gain and PDE. A prolonged cryogenic test at higher OV was deemed necessary to demonstrate response stability at these elevated gain levels. Consequently, we developed a cryogenic readout system to monitor DUNE SiPM arrays operating at 3 V, 4 V, and 5 V in liquid nitrogen ($LN_2$) for over three months. Key parameters of dark rate, energy resolution, signal-to-noise ratio (SNR), total correlated noise, charge gain, and photodetection efficiency (PDE) - were continuously tracked. Our primary objective is to detect any drift beyond statistical fluctuations rather than to measure absolute values of the parameters.

The photoelectron signals from 54 SiPMs (in 3 groups of 18 SiPMs each, biased at 3 V, 4 V and 5 V OV) were read out simultaneously in $LN_2$ using two µFEMB boards (micro-Front-End Mother Board), recently developed for noble liquid detector readout R&D [2-4]. Each µFEMB integrates two cryogenic programmable front-end charge-sensitive amplifier ASICs (Application-Specific Integrated Circuits), known as LArASICs, providing 32 signal input channels [5,6]. We implemented a waveform snippet acquisition scheme, capturing 70 µs per waveform (140 samples at a 2 MS/s sampling rate) whenever the signal amplitude exceeded a predetermined trigger level. Empty waveforms were discarded to optimize data throughput and storage. A 450 nm LED, pulsed at 100 Hz and fiber-coupled to the SiPM array, provided quasi-uniform illumination for PDE measurements across all SiPMs. Data acquisition and first-level analysis were fully automated and synchronized.

### 1.2 SiPM Devices

The SiPM model used is the HPK S13360-6075-HS-HRQ (S13360-9935). It has an effective photosensitive area of 6×6 mm², containing 6364 pixels with a 75 µm pitch, a hole-wire-bond ball anode at the center, and a 150 µm thick hard silicone resin entrance window [Figs. 1(a)-(c)]. Six SiPMs are mounted on a PCB to form a subset of a photodetection module, with cathodes tied together to form a common cathode [Fig. 1(d)]. The device features High Quenching Resistance (HRQ) specifically designed for use in the FD1 module, in accordance with DUNE Physics Requirements [7]. From room temperature to liquid nitrogen temperature, the measured total HRQ increases from ~89 Ω to ~320 Ω (0.56 MΩ to 2 MΩ per pixel), while the terminal capacitance drops from ~1.88 nF to ~0.89 nF. This increases the overall recovery time of the SiPM from ~0.19 µs to ~0.65 µs, [Figs. 2(a)-2(b)], which remains well within the 1 µs peaking time of the LArASIC, thus avoiding ballistic deficit.

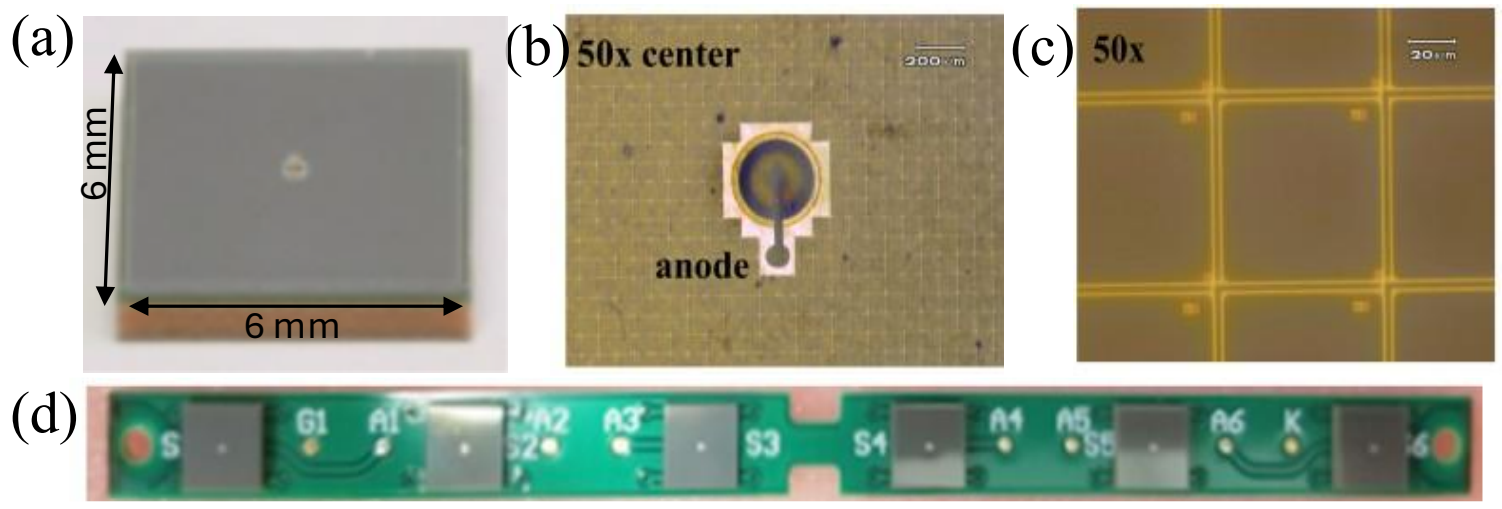

**Figure 1.** (a) HPK S13360-9935 SiPMs with an effective area of 6x6 $mm^2$, (b) 50x magnification showing the ball anode at the center, (c) higher magnification showing the 75μm pixel pitch structure, (d) Six SiPMs on a PCB formed a subset of DUNE module with a common cathode.

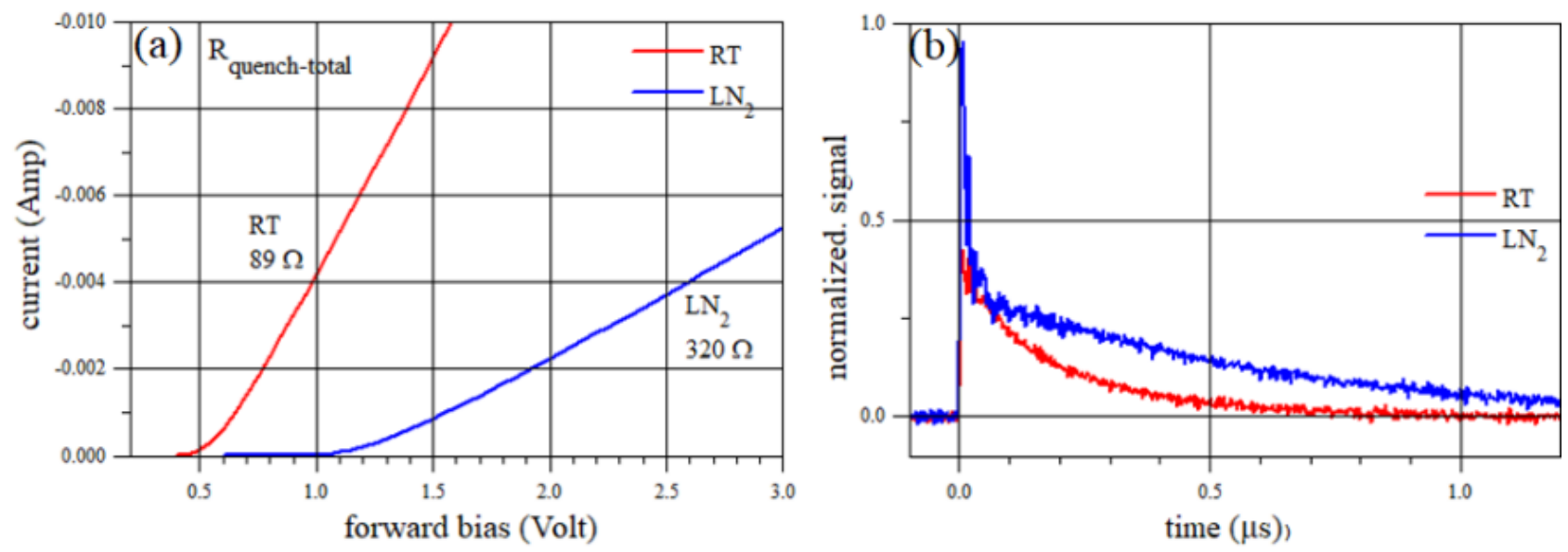

**Figure 2.** (a) Forward biased IV of HPK S13360-9935 SiPMs at room and $LN_2$ temperatures, (b) corresponding direct SiPM pulse shape.

The room-temperature I-V characteristics of all 54 SiPMs were measured beforehand, while the $LN_2$ I-V curves were measured after the experiment. Breakdown voltages ($V_{breakdown}$) were determined by Gaussian fitting of the first derivative of the log(I) vs. $V_{bias}$ plots [Figs. 3(b) and 3(e)]. In $LN_2$, the extremely low thermal dark current necessitates photon injection to establish a clear breakdown voltage. A decrease of five orders of magnitude in dark current from room temperature to $LN_2$ was observed for all devices. The statistical averages of $V_{breakdown}$ shown in Figs. 3(c) and 3(f), are ~51.71 V (σ ~0.05 V) at room temperature and ~41.86 V (σ ~0.07 V) in $LN_2$. These small $V_{breakdown}$ variations correspond to intrinsic charge gain variations of ~1.25% and ~1.75% at 4 V OV for all SiPMs at room temperature and in $LN_2$, respectively.

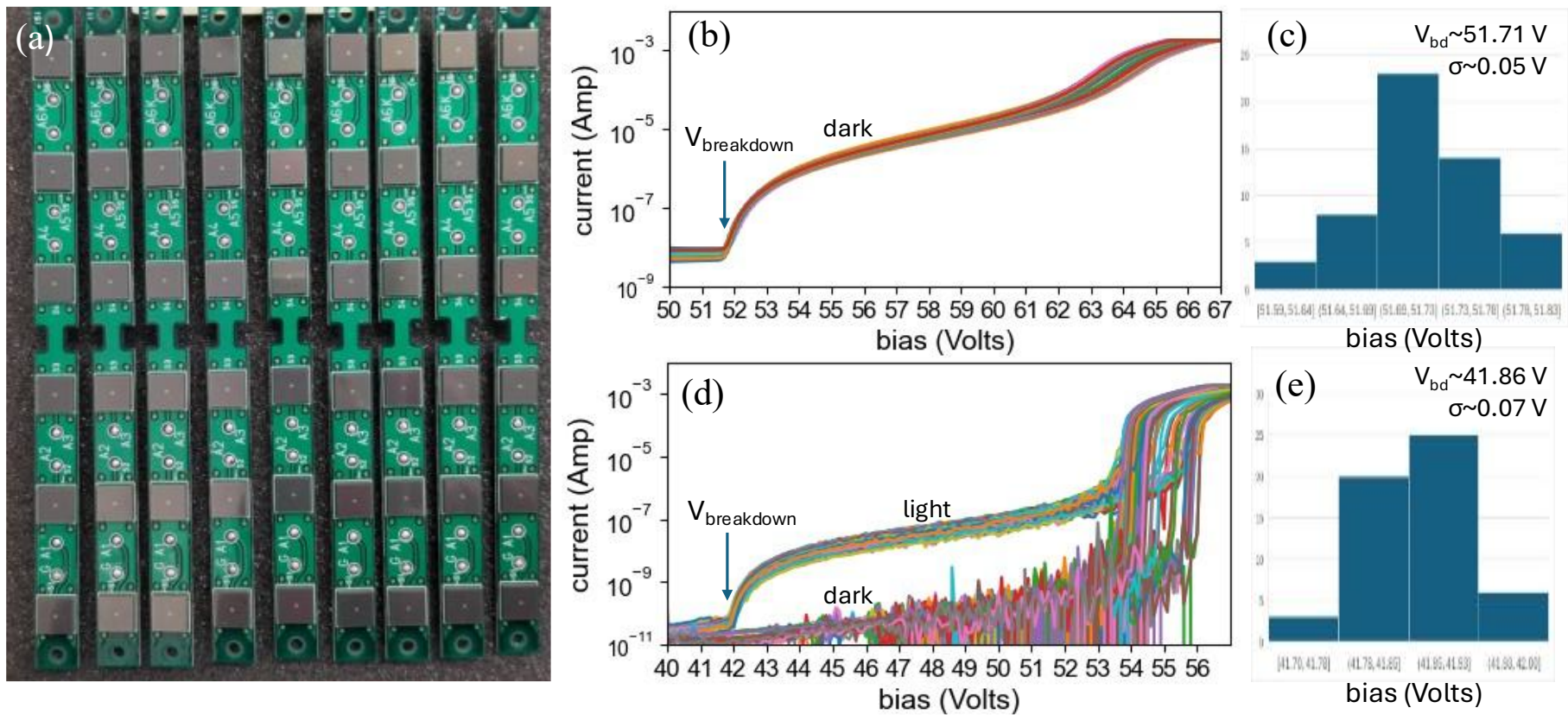

**Figure 3.** (a) 6x9 SiPMs mounted on PCB broads formed subsets of DUNE module, (b-c) IV and histogram of the $V_{break-down}$ for all 54 SiPMs at room temperature, and (d-e) in $LN_2$.

## 2. Experimental

Our approach for reading out charge from SiPMs with large intrinsic capacitance is by weak capacitive coupling to a charge-sensitive amplifier. When there is a large mismatch between the detector capacitance $C_d$~nF and the amplifier's input transistor gate capacitance $C_g$~fF, an optimal solution for efficient charge transfer is to use a small decoupling capacitor $C_b$ where $C_d > C_b >> C_g$. This configuration preserves the equivalent noise charge (ENC) of the readout system; a detailed analysis of this concept is available in recent literatures [8-10]. Consequently, each SiPM signal waveform is coupled to the corresponding front-end readout channel via an independent 10 pF decoupling capacitor to optimize the dynamic range.

### 2.1 Readout Electronics

The μFEMB integrates 2 LArASIC chips (P5B version), 32 single-channel 12-bit Commercial-Off-The-Shelf (COTS) ADCs, an Altera Cyclone IV FPGA to converted 32-channel charge signals into two 1.25 Gb/s high-speed links over coaxial cables in the cryogenic environment. Two μFEMB boards (using 30 and 24 channels each) collected all 54 SiPM signal waveforms. Each LArASIC contains an internal 185 fF calibration capacitor and an on-chip 6-bit DAC (amplitude adjustable from 0 to 1.2V) used as pulser for charge calibration. Additionally, two 1 pF capacitors were connected to channels 0 and 49, respectively, for system-wide charge calibration, yielding ~1 pC ≈ 190 ADC counts. The 10 remaining unconnected channels measured an intrinsic system electronic noise as ~0.5 ADC counts (~290 RMS electrons), compared to ~0.64 ADC counts on connected channels increased due to the ~nF large SiPM capacitance.

A firmware upgrade on the μFEMB board enabled the waveform snippet acquisition scheme. While the ASIC charge gain and shaping time are dynamically programmable (gain: 4.7, 7.8, 14, and 25 mV/fC; peaking time: 0.5, 1, 2, 3 μs), they were fixed at 4.7 mV/fC and 1 μs, respectively. The 2 MS/s ADC sampling rate is locked to a 10 MHz reference clock. A total of 140 samples or 70 μs per waveform (0.5 μs/time-tick) was collected with 10 ns timestamp accuracy, satisfying the Shannon-Nyquist sampling theorem for digital reconstruction. Waveform recording triggered whenever the signal amplitude exceeded ~0.3 photoelectrons (~0.3-p.e.) for dark count, or when 100 Hz LED trigger received. Empty waveforms were discarded to optimize data throughput and storage. Data was streamed via Python through an optical transceiver for real-time analysis, with detailed processing performed offline. Two uninterruptible power supplies (UPS) safeguarded the system over the three-month test period.

The 54 SiPMs were divided into three groups of 18. Negative bias voltages of 45, 46, and 47 V were applied to the anodes, resulting in OV of 3, 4, and 5 V (designated HV1, HV2, and HV3, respectively). Waveforms are negative-going, with baselines at high ADC levels and signal amplitudes at low ADC levels. A 1-MΩ series resistor limited the high-voltage ripple noise and the current draw for each power supply, which were remained at the instrument limit (~nA).

### 2.2 Photon Illumination

For photon delivery (PDE monitoring), a 465 nm blue LED was pulsed to ~10 ns duration at a 100 Hz repetition rate [see inset of Fig. 4(a)]. The 465 nm wavelength corresponds to the peak of the spectral response of the DUNE SiPMs. Light was delivered via a 3 m long optical fiber (Thorlabs FG600AEA) butt-coupled through an SMA connector to a 3.2 mm diameter glass diffuser (Edmund Optics #38-301, 152.4 mm long image conduit with 12 μm fibers,). This configuration provided quasi-uniform illumination across all SiPMs in a cylindrical geometry for PDE monitoring, Fig. 4(a). Light intensity was adjusted to the single-photon level for all devices.

The SiPM modules and the photon delivery system were oriented vertically, with diffusely scattered photons radiating cylindrically to SiPMs at varying heights. The bottom exit of the glass diffuser was capped with a black SMA fiber termination connector. The uniformity of the diffusely scattered photons, evaluated at ambient temperature, was within a factor of two. This vertical positioning minimizes the deposition of microscopic debris on SiPM surfaces and reduces terrestrial cosmic muon interactions with the SiPM entrance windows, thereby eliminating the event burst phenomenon [11,12].

The cylindrical SiPM test stand was first made light-tight with black aluminum foil, placed in a basket [Figs. 4(b) and 4(c)], and then submerged in $LN_2$ in a dedicated 1 m deep, light-tight, non-air-tight dewar, Fig. 4(d). Six thermocouples were strategically positioned inside the dewar at various heights to monitor the $LN_2$ level and signal the need for refilling every 4 to 6 days. All data acquisition was performed hours after $LN_2$ replenishment to ensure a steady-state thermal environment.

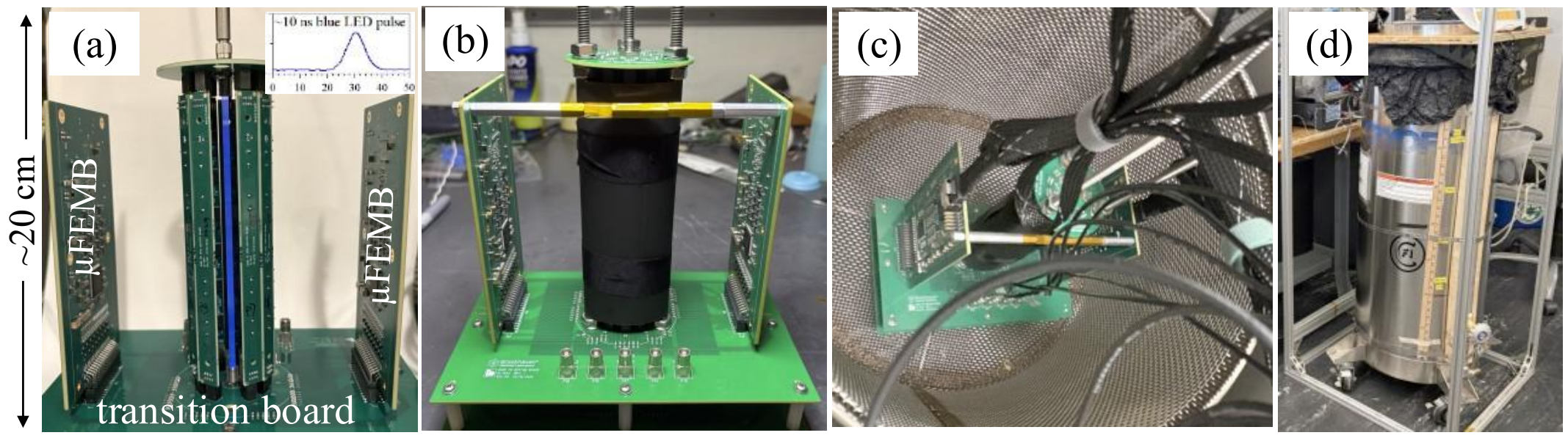

**Figure 4**. The cylindrical SiPM test stand with blue LED illumination (a) before and (b) after light-tight with aluminum foils, (c) after electrical and optical connections then placed in a basket, and (d) deployed in the dedicated dewar.

### 2.3 Waveform Collection

The typical thermally activated random dark rate of SiPM in $LN_2$ is <1 Hz, which would require a prolong collection time to build up sufficient statistics. Therefore, all dark pulses with amplitude >0.3 pe (the trigger threshold) were recorded with a window of 20 µs (40 samples) before and 40 µs (80 samples) after the trigger. The total dark pulse count per channel was recorded every 5 minutes.

We noted that some electronic devices operating in the vicinity of the test-stand can emit RF transients, temporarily affecting the Dark Count Rate (DCR). Additionally, the passage of cosmic rays in $LN_2$ would generates Cherenkov light seen by the SiPMs which elevates the DCR. Consequently, the DCR was reported with a 20 Hz cut applied to the dark rate frequency distribution. The statistical mean of the dark pulse rate, ~2 Hz per SiPM (~0.05 Hz/mm$^2$), was reported every 5 minutes, The single-photoelectron (SPE) histogram for each channel was generated and visually displayed hourly, with the statistical fluctuation $\sigma$ defined here as the variation of the mean value per hour.

Representative subsets of time-unfolded and time-folded random dark and LED-triggered waveforms are shown in Figs. 5(a)-5(f). Dark pulses were continuously acquired at a 0.3-p.e. trigger threshold, resulting in ~$10^6$ waveforms/day/device, while LED illumination was programmed to run three times per day (at 00:00, 08:00, and 21:00) for five minutes each, resulting in ~$10^5$ waveforms/day/device.

The random nature of thermal dark pulses results in a time dithering of the signal peak by 1-time-tick, Fig. 5(e), while the dithering is completely absent in the 10 MHz time-locked LED light pulses, Fig. 5(f). SPE histograms were computed from the dark and LED peak signal amplitudes, Figs. 5(g) and 5(h). Typically, more than five photoelectron peaks were observed on the dark pulses. For LED pulses, more than seven photoelectron peaks were observed, including the 0th photoelectron peak, where a trigger was received but no photon arrived, Fig. 5(b). Long tail of after-pulses caused by charge carriers trapped in silicon defects and then released later in time, were captured following the main pulse, Figs 5(c) and 5(d). These were not analyzed here, as they typically contribute <2% of the total correlated noise. However, after-pulses occurring within the 1 μs peaking time of shaping amplifier were collectively integrated into the main pulse and appeared on the right shoulder of every photoelectron peak; these were analyzed as correlated noise, Figs. 5(g)–5(h). We note that that 0-p.e. peak in Fig. 5(h) contains only electronic noise (no photons and exhibits no such shoulder – a hallmark of SiPM SPE histograms.

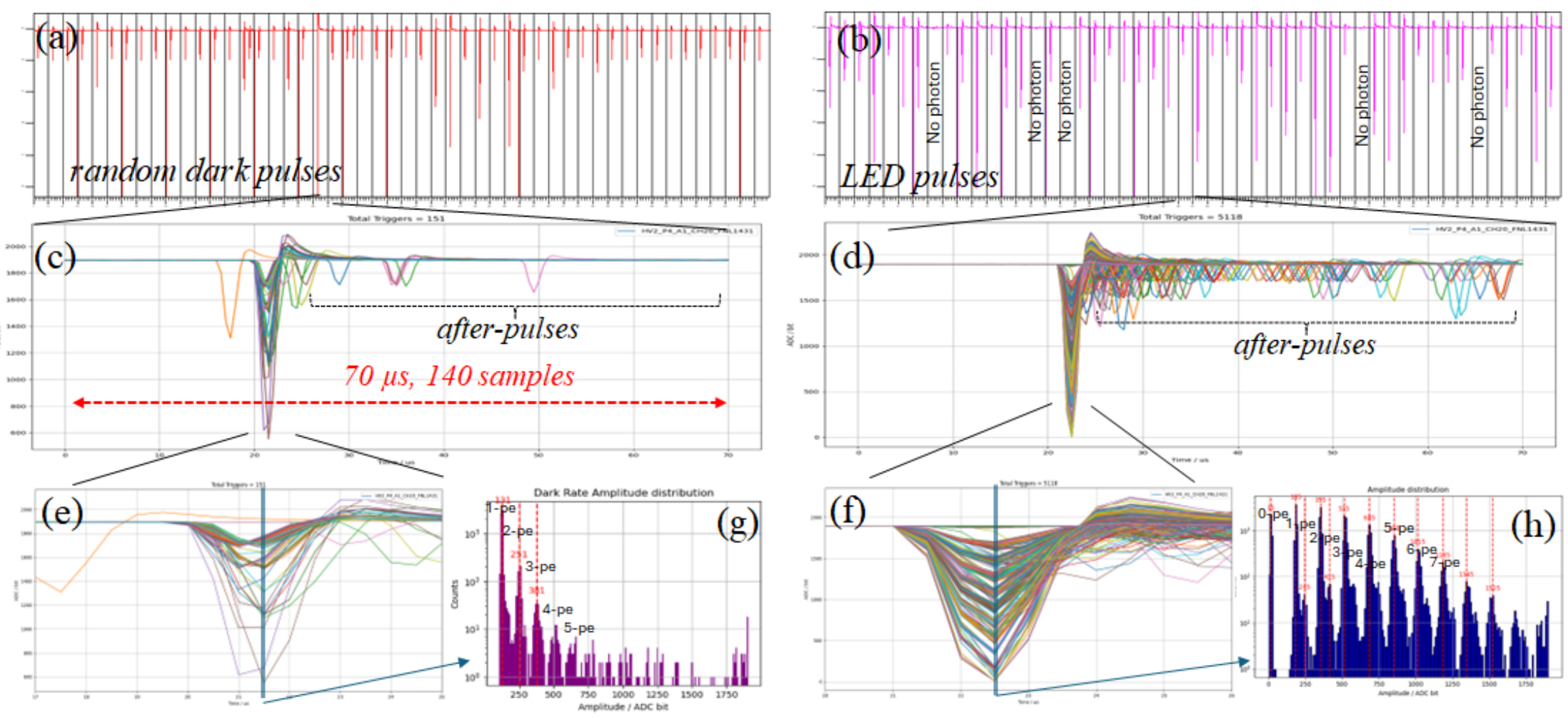

**Figure 5.** Subsets of (a)(b) time-unfolded and (c)(d) time-folded dark and LED waveform traces, and (e)(f) the expanded view on the amplitude of their photoelectron peaks, and (g)(h) the representative SPE histograms displayed in log count scale.

## 3. Results

The SiPM parameters monitored in this study include: (1) dark count rate, (2) energy resolution, (3) signal-to-noise ratio, (4) total correlated noise, (5) charge gain, and (6) photodetection efficiency. We are not targeting the absolute values of the parameters, only aim to detect drifts, if any, beyond statistical fluctuations. Therefore, all measured values are normalized to the 3-month average value of each respective SiPM. Parameters were derived from the SPE histograms of the dark and LED pulses for every device; representative histograms are shown in Fig. 6. We defined the normalized 1-p.e. resolution as $\frac{\delta E_{1pe}}{\Delta E} / \left\langle \frac{\delta E_{1pe}}{\Delta E} \right\rangle$, the SNR as $\frac{X_{1pe}-X_{0pe}}{\delta E_{0pe}} / \left\langle \frac{X_{1pe}-X_{0pe}}{\delta E_{0pe}} \right\rangle$, and the charge gain as $\frac{\Delta E}{\langle \Delta E \rangle} = \frac{X_{2pe}-X_{1pe}}{\langle X_{2pe}-X_{1pe} \rangle}$. In these expressions, $\Delta E$ represents the spacing between the photoelectron peaks, $X_{0pe}$, $X_{1pe}$, and $X_{2pe}$ are the amplitudes of the 0, 1, and 2-p.e. peaks, and $\delta E_{0pe}$ and $\delta E_{1pe}$ the fitted Gaussian width of the 0-p.e. and 1-p.e. distributions. The notation ‹…› denotes the average of the corresponding values over the 3-month period, see Figs. 6(a) and 6(b).

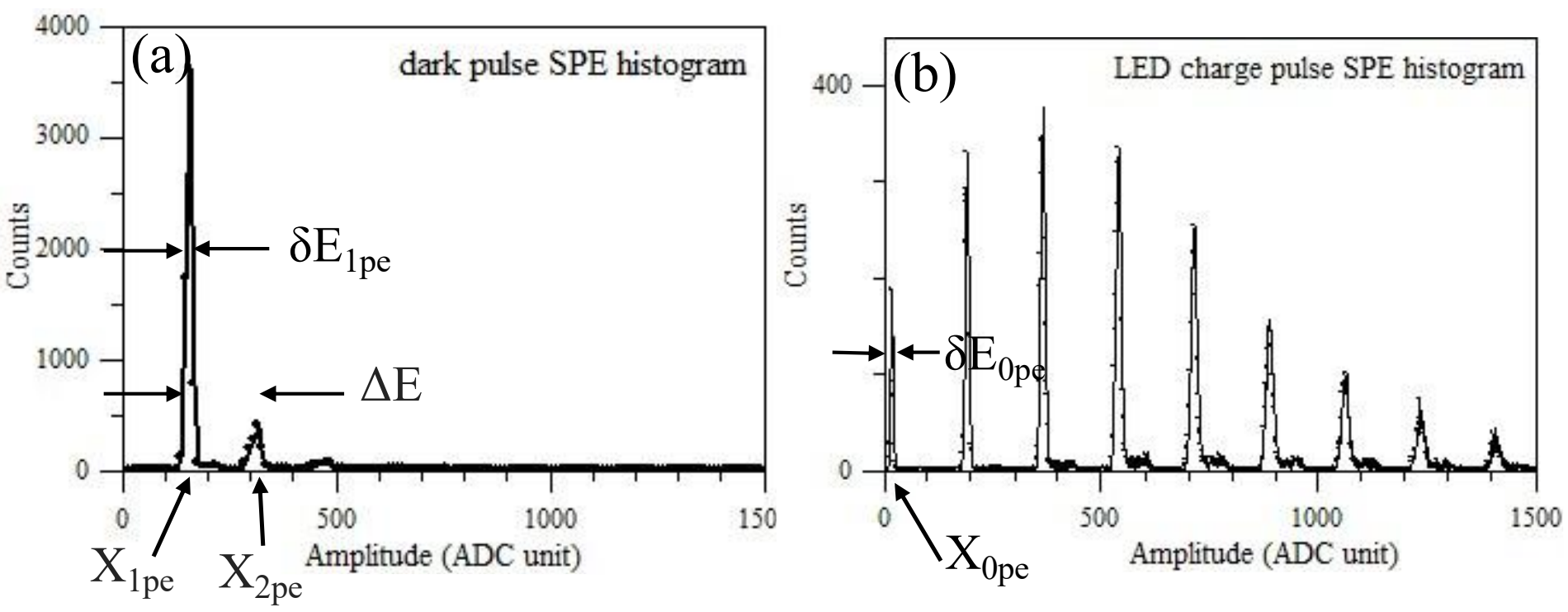

**Figure 6**. Representative (a) dark pulse and (b) LED pulse SPE histograms. $X_{0pe}$, $X_{1pe}$, and $X_{2pe}$ are the amplitudes of the 0, 1, and 2-photoelectron peaks. $\Delta E$ is the spacing between 1- and 2-photoelectron peaks, $\delta E_{0pe}$ and $\delta E_{1pe}$ are the fitted Gaussian width of the 0- and 1-photoelectron peaks from the LED and dark histograms, respectively.

### 3.1 Dark Count Rate (DCR)

Dark counts are spurious avalanche discharge triggered by thermally or field-generated electron–hole pairs rather than photons. The DCR of each SiPM varies slightly but remains within a few percent. It increases nominally with OV, with a statistical average of 1.52, 1.81, and 2.12 Hz from 3, 4, to 5V OV, respectively, across all 54 devices. These rates amount to ~50 mHz/mm$^2$ at 4V OV for the 75 µm pixels. Figures 7(a)-(c) show representative DCR histograms, while Figs. 7(d)-(f) display the corresponding normalized DCR over the 110-day period. Some DCR outliers were observed, which may be attributed to fluctuations in environmental electronic noise. No noticeable difference in dark pulse amplitude histogram was observed over the three-months test period. However, there is a slight but gradual drop in DCR, exceeding the ~1σ statistical fluctuation. Figure

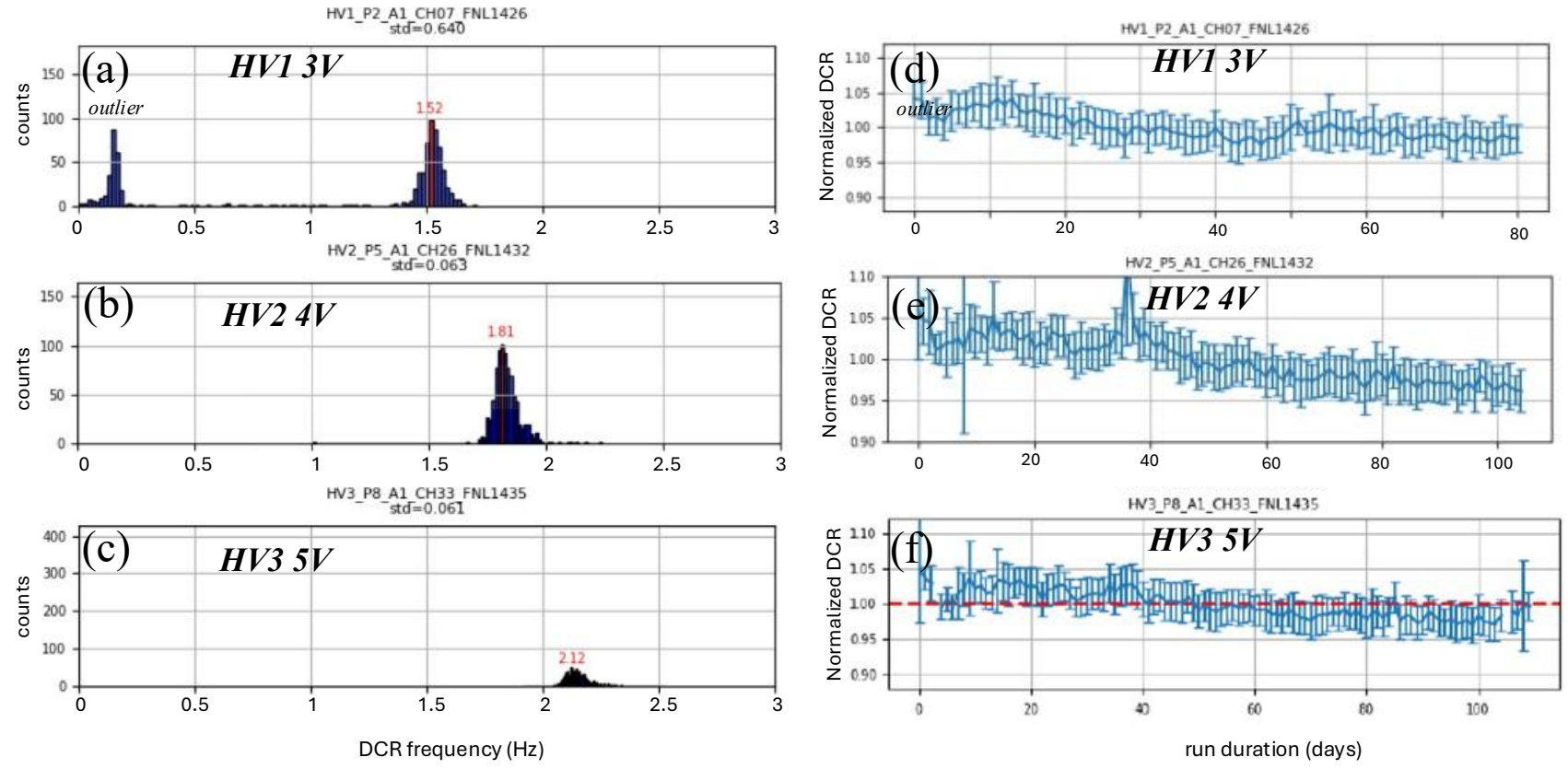

**Figure 7**. (a-c) Representative hourly DCR histograms at 3, 4 and 5 V OV, and (d-f) their corresponding normalized DCR over run days, 1σ error bars are the statistical variation of DCR over that day.

8 displays the normalized DCR over 110 run days for the 18 SiPMs biased at 4 V OV; similar drop was observed for all other devices. Notably, there is no evidence of this drift being enhanced at 4 V, and 5 V compared to 3 V OV. Since the electronic noise remained stable for all devices, we speculated that ~1σ drop in DCR may result from charge carriers being progressively trapped in silicon defects over time during continuous cryogenic cooling.

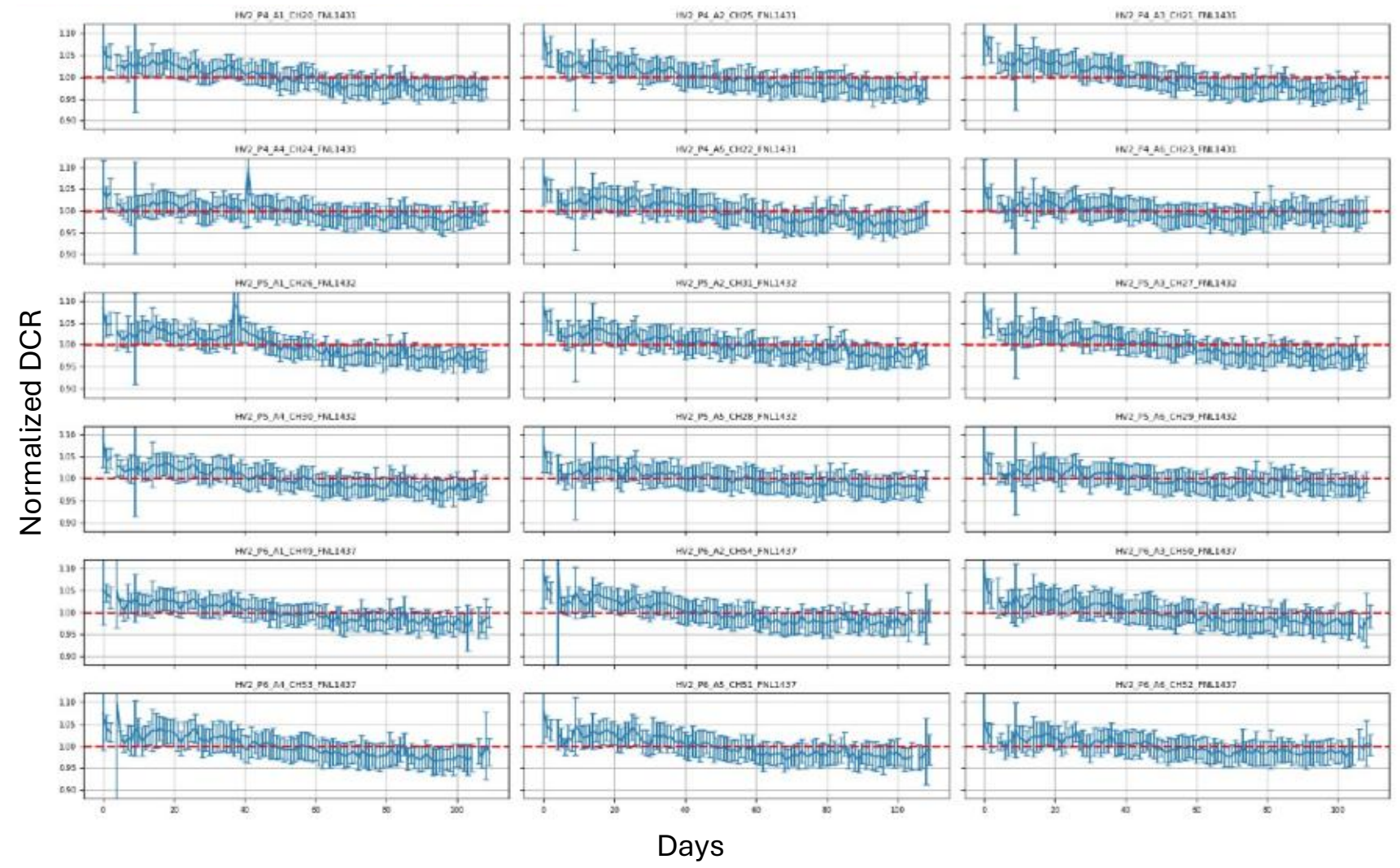

**Figure 8**. A selected set of normalized DCR over 110 run days for all 18 SiPMs biased at 4V OV.

### 3.2 One-Photoelectron Resolution

A hallmark characteristic of SiPMs is their excellent Photon Number Resolving (PNR) capability, which improves significantly at cryogenic temperatures due to the extremely low dark count rate. The resolution of the photoelectron peaks directly impacts the energy reconstruction of a photon detector, thereby affecting the sensitivity of physics explorations. Generally, a SiPM with lower terminal capacitance yields higher photoelectron resolution and superior PNR power. Figure 9 shows the normalized the 1-p.e. resolution over the test period for the selected 18 SiPMs at 4 V

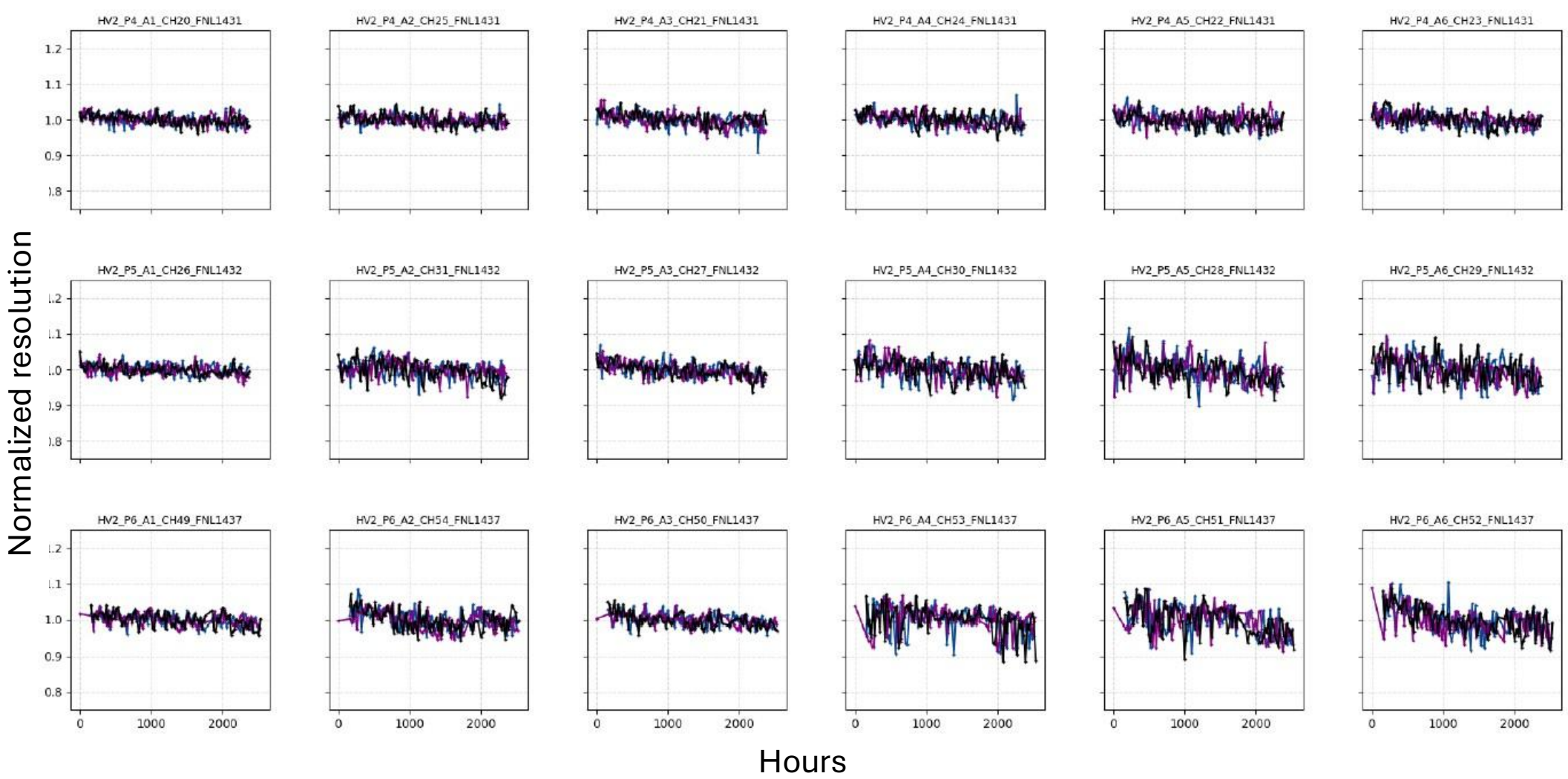

**Figure 9**. A selected set of normalized 1-p.e. resolution over 110 run days for all 18 SiPMs biased at 4V OV.

OV; similar results were observed for all other devices. We found that the 1-p.e. resolution of random dark pulses fluctuates <5% on most channels and <10% on others. A slight ~1σ drop in 1-p.e. resolution was noticed for a few devices, but there was no evidence of drift enhancement at higher OV.

### 3.3 Signal-to-Noise Ratio (SNR)

As stated earlier, the noise of a typical SiPM channel is ~0.65 ADC counts, while the signal amplitude varies from 150 to 230 ADC counts across channels at different bias OV. A high SNR of >100 is inferred, primarily due to the low SiPM terminal capacitance and the excellent low-noise characteristics of the μFEMB. Figure 10 shows the normalized SNR over the test period for the 18 SiPMs at 4 V OV. We observed a slight increase in 1-p.e. SNR over time, suggesting that the electronic noise of the system decreases slightly (as the 1-p.e. amplitude, or charge gain, remains constant, as shown in the next section). Similar increases in SNR were observed on all other devices. Some outliers occurred due to unfavorable Gaussian fits of the 0-p.e. peak on certain channels. Again, no evidence suggests this drift is enhanced at 4 V or 5 V compared to 3 V OV.

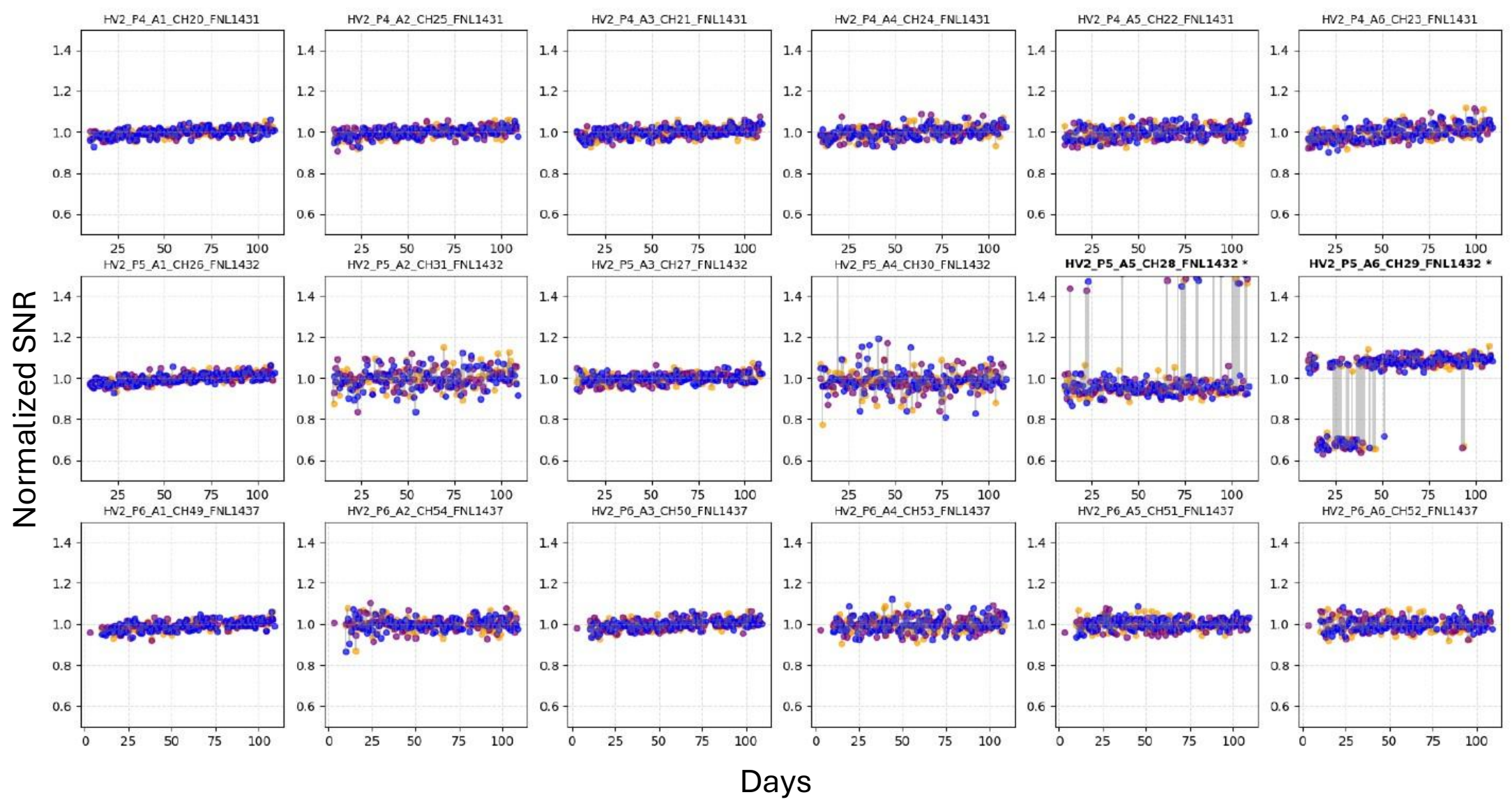

**Figure 10**. A selected set of normalized SNR over 110 run days for all 18 SiPMs biased at 4V OV.

### 3.4 Correlated Noise (CN)

Correlated noise consists of secondary avalanche events in SiPMs that are stochastically triggered by a primary signal photon or a dark pulse. Unlike uncorrelated dark noise, which occurs randomly, correlated noise is physically linked to a preceding discharge and can significantly impact the energy resolution and timing of a photon detection system. Because CN only appears in correlation with genuine thermal or photon events (1-p.e., 2-p.e., etc.), it manifests as an increase in the mean charge of those events. The CN is measured by subtracting the unweighted count from the normalized 1-p.e. weighted count based on the integral of the dark histogram. We calculate the correlated noise as: $\mathrm{CN(pe)} = \frac{\sum_{i=1}^{N}(X_i Y_i)/X_{1pe}}{\sum_{i=1}^{N} Y_i} - 1$, where $X_i$ is the amplitude value in ADC units, and $Y_i$ the corresponding count rate, summing over all bins N within the range 0.5 p.e.$<N<$4.5 p.e. [13]. The total correlated noise examined here, CN(p.e.) represents the average number of additional photoelectrons per primary discharge. Thereby, CN(p.e.) is a pedestal event counter. This includes crosstalk ($<$4.5 p.e.) and after-pulses occurring within the 1 μs peaking time of the shaping amplifier, see Fig. 6(a).

The total CN for all 54 SiPMs biased at 3, 4, and 5 V OV is shown in Fig. 11, where the average values and error bars for each device are presented. The average CN for each group is shown in the inset of Fig. 11. Normalized CN values for the selected 18 SiPMs biased at 4 V OV throughout the 110-day period are presented in Fig. 12. It is evident that CN increases nominally with OV; however, an apparent ~1σ increase in CN was observed over the three-month test period. A similar increase was observed across all other devices. Nevertheless, there is no evidence of this drift being enhanced at 4 V or 5V compared to 3 V OV. It is plausible that this small increase in CN may result from excess external crosstalk enhanced by the accumulative particulates in $LN_2$. Since optical crosstalk is caused by photon emission during the primary avalanche that triggers secondary discharge in adjacent cells, the emitted (dark) photons may back-scatter by the particulates to trigger the neighbor cells.

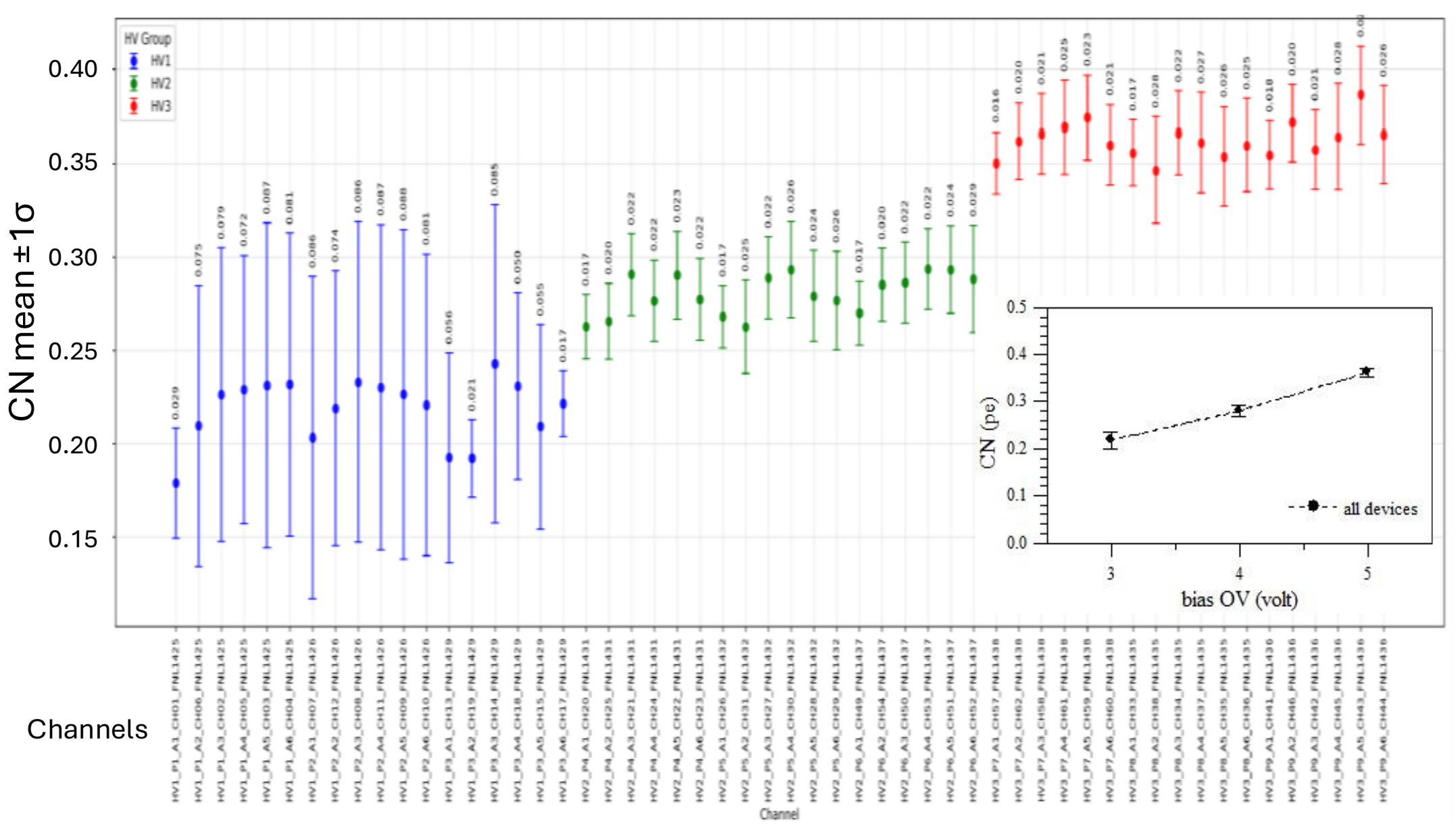

**Figure 11.** The total CN(pe) for all 54 SiPMs biased at 3, 4, and 5V OV. Blue, green, and red dots and bars are, respectively, the average value and the error bars for each device. Inset is the average CN(pe) for each group.

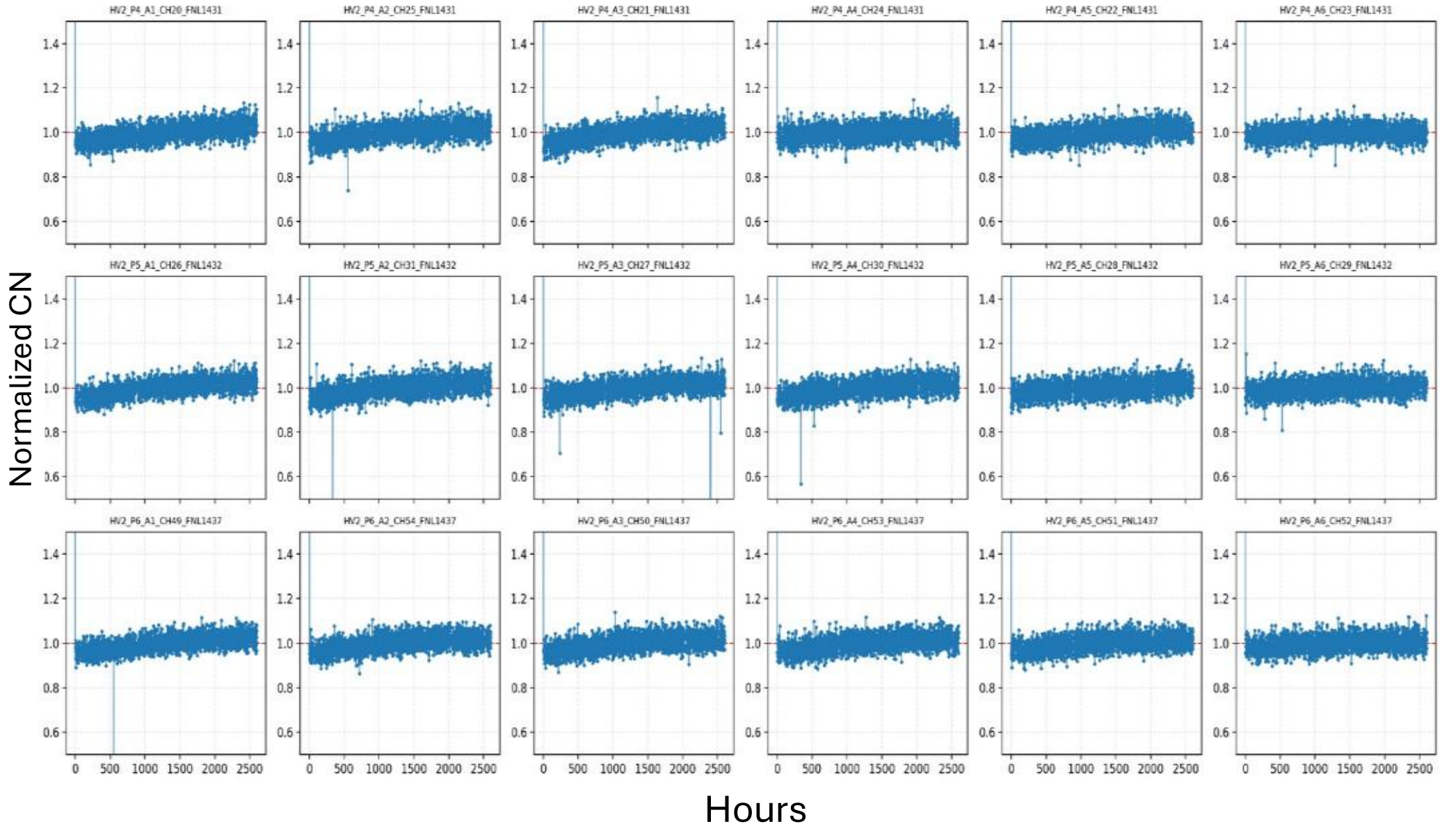

**Figure 12**. A selected set of normalized correlated noise over 110 run days for all 18 SiPMs biased at 4V OV.

## 3.5 Charge Gain

SiPMs are high-gain photon detectors that operate in Geiger mode, where a single photon triggers an avalanche multiplication producing $10^5$ to $10^7$ electrons. This gain is linearly proportional to the OV and microcell capacitance. In a typical SPE histogram, the 1-p.e. charge gain is proportional to the spacing between photoelectron peaks, $\Delta E = X_{2pe} - X_{1pe}$, see Fig. 6(a). The charge gain of each SiPM varies slightly due to the small $V_{breakdown}$ variations discussed in Section 1.3;

however, these fluctuations remain within a few percent across all bias levels. Based on the charge calibration described in Section 2.1, the absolute charge gain measured here agrees with specifications, ~$6 \times 10^6$ at 4 V OV.

No charge gain drift was observed in dark pulses acquired at 3, 4, or 5 V OV. A selected dataset is shown on Fig. 13. An apparent splitting of charge gains into two bands at 5 V bias, visible in Fig. 13(c), is an artifact resulting from the time dithering of the signal peak by one sampling period (one time-tick) due to the random arrival times of thermal dark pulses.

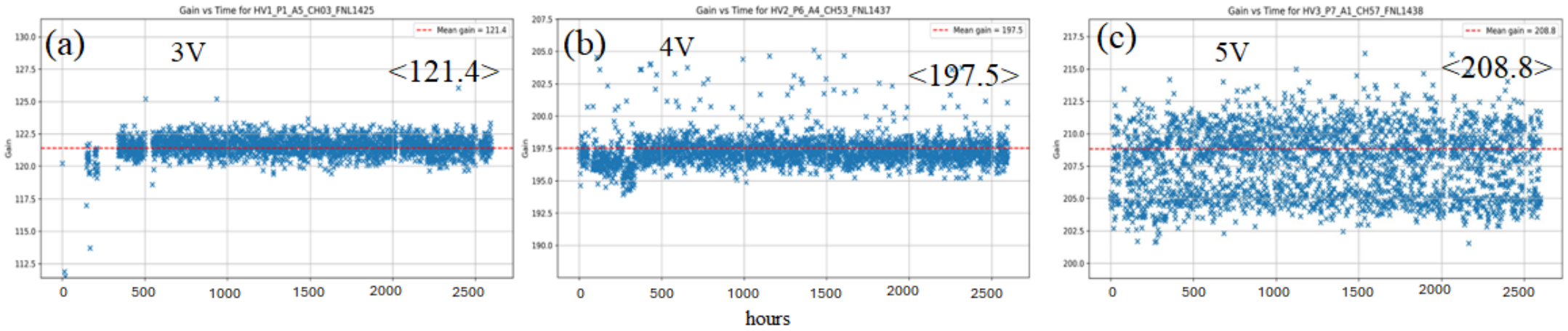

**Figure 13**. A selected set of 1-p.e. charge gain, which is proportional to the spacing between the photoelectron peaks of the dark pulse for all 3 SiPMs biased at (a) 3, (b) 4, and (c) 5V OV over 110 run days, the statistic average of each channel is highlighted in ‹…›.

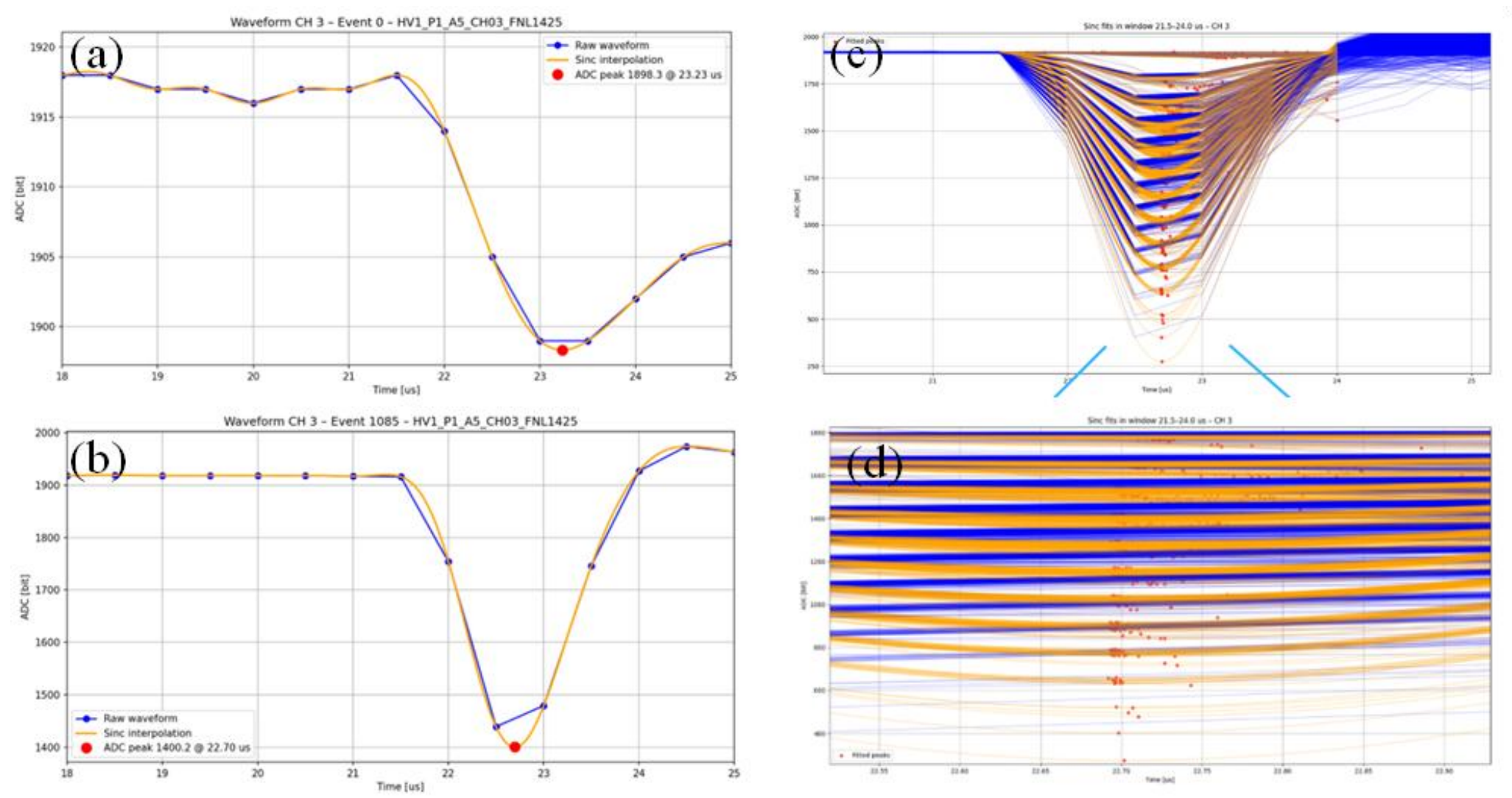

**Figure** 14. A representative LED waveform analysis where a sinc-interpolated fitting window that varies daily for each channel was employed to restore the true peak amplitude of each waveform, (a) 1-p.e. waveform, (b) 4-p.e. waveform, (c) all photoelectron peaks, (d) zoom in on the peaks of the waveforms, red dots depict the sinc-interpolated ADC peaks.

When examining the charge gain using synchronized LED light pulse data, we initially observed a gain oscillation with a periodicity of ~11 days, this was caused by a long-term phase drift of the 10-MHz reference clock. To correct this, rather than using a simple peak-finding algorithm, we applied a combined sinc-interpolation and peak-finding algorithm to fit every waveform prior to analyzing charge gain and PDE. Figure 14 illustrates the LED waveform analysis procedure, where a channel specific fitting window was employed daily to restore the true peak amplitude. Consequently, we confirmed there no physical charge gain drift occurred over the 110-day test period at any OV, although a residual ~1% oscillation remained across all devices, Fig. 15. Notably, no such oscillation was observed when the 10 MHz lock was disabled and waveforms were triggered by 0.3-p.e. threshold instead. Furthermore, there was no evidence of enhanced charge gain drift or variation when operating at 4 V or 5 V compared to 3 V OV.

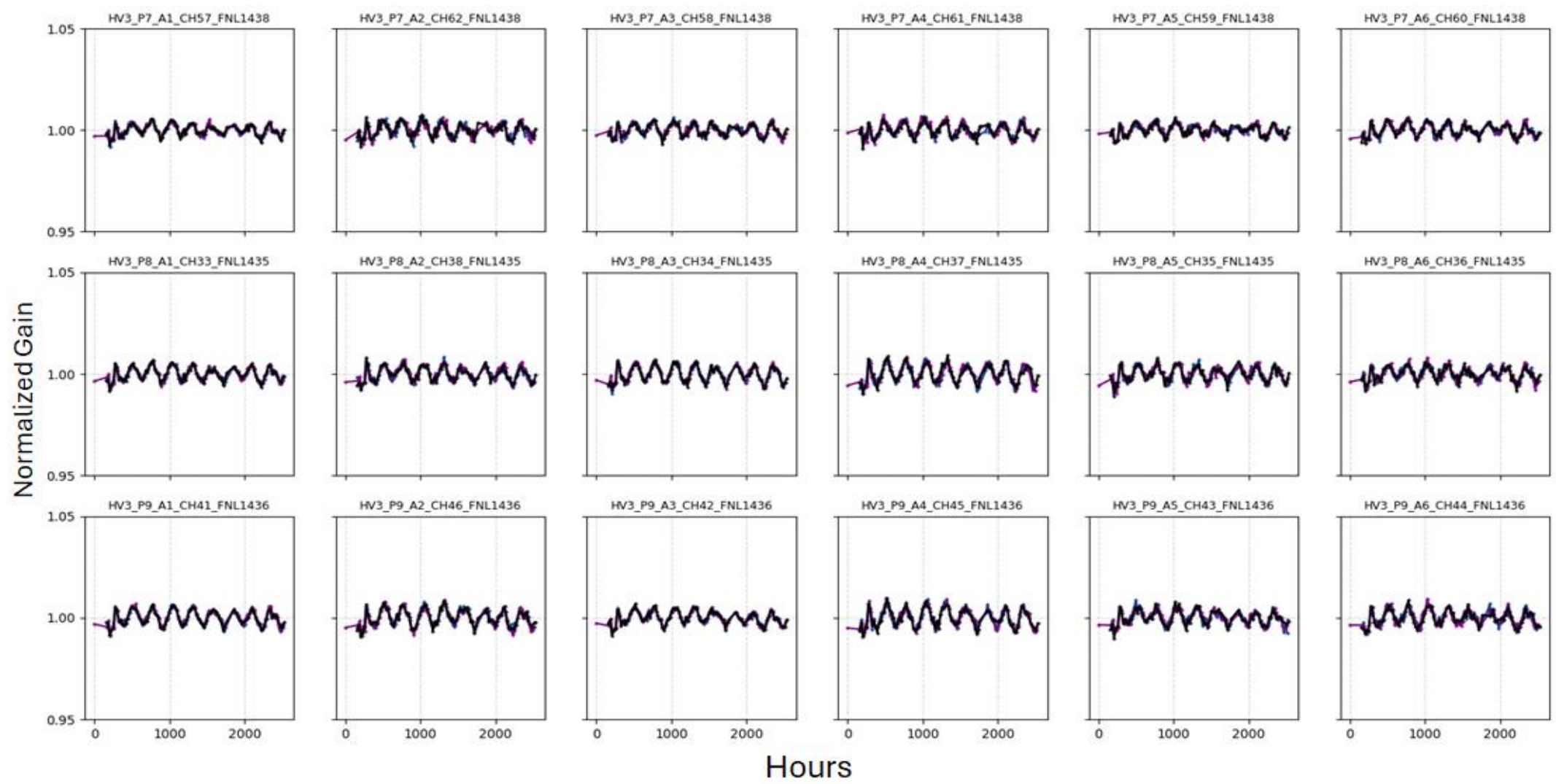

**Figure 15**. A selected set of normalized charge gain from LED light pulse for all 18 SiPMs biased at 5 V OV over 110 days. A residual ~1% charge gain oscillation with a periodicity of ~11-days on all devices due to the imperfect long-term phase drift of the 10 MHz phase-locked reference clock.

### 3.6 Photo Detection Efficiency (PDE)

Photon Detection Efficiency (PDE) is a critical metric for SiPMs, representing the probability that an incident photon will trigger a detectable output signal. PDE is the product of three independent factors: quantum efficiency, Geiger-mode triggering probability, and the geometric fill factor of the device. To monitor the PDE, we utilized a 10 MHz time-locked, 450 nm LED with 10 ns pulse, fiber-coupled via a glass diffusor bundle. The mean number of photoelectrons is associated with the PDE and can be extracted from a typical SPE histogram using either a Gaussian envelope fit or Poisson statistics.

Alternatively, the Poisson method calculates the mean, $N_{pe}$(Poisson), using the zero-photoelectron (pedestal) data: $\mathrm{N_{pe}(Poisson)} = -\ln\left(\frac{\mathrm{N_{pedestal}}}{\mathrm{N_{total}}}\right)_{LED} + \ln\left(\frac{\mathrm{N_{pedestal}}}{\mathrm{N_{total}}}\right)_{Dark}$, where $N_{pedestal}$ , $N_{total}$ represent the number of events in the 0-p.e. peak and the total number of events, respectively. The first term is computed from the LED SPE histogram, while the second term accounts for the dark pulses from the same device.

Figure 16 shows representative LED SPE histograms at (a) 3 V, (b) 4 V, and (c) 5 V OV before (black) and after (red) the three-month run period, along with their Gaussian envelope fits. The mean values deduced from these fits are highlighted in the insets. It is evident that the mean number of photoelectrons decreased over the three-month period. Figure 17 presents a selected set of normalized PDE values, calculated via the Poisson method, for the 18 SiPMs biased at 5 V. A similar drop in PDE was observed across all other devices.

Because no charge gain drift was observed - indicated by the constant spacing between SPE peaks - and because we independently confirmed that the LED light delivery system stability was ~0.5% (RMS) with no long-term intensity degradation, we infer that the observed PDE drop is physical. This may be attributed to cumulative optical contamination of the liquid nitrogen impeding photon transmission or a physical aging effect within the devices, both of which warrant further investigation.

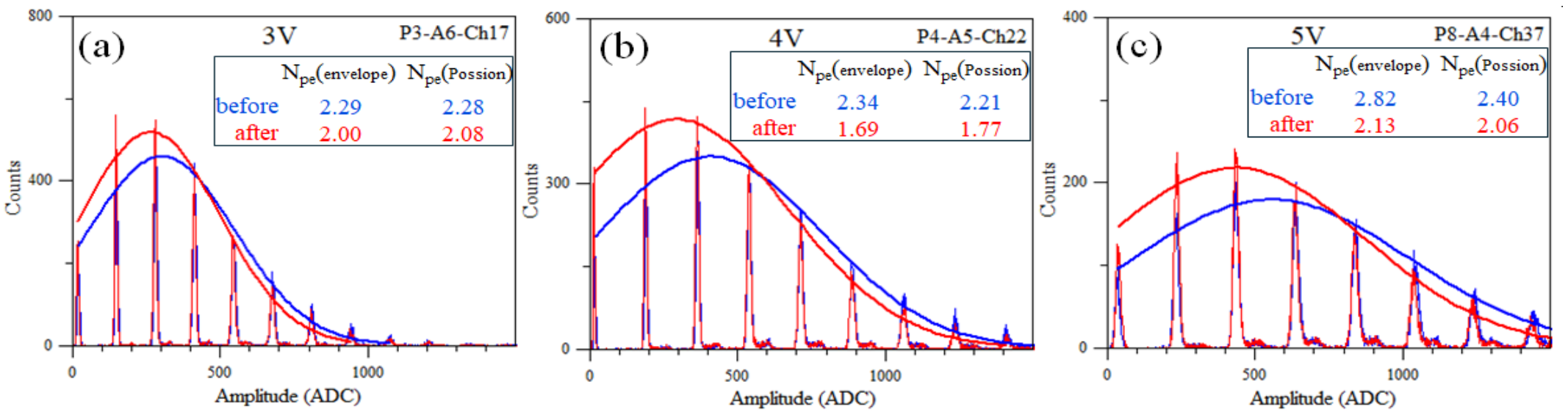

**Figure 16**. Representative LED SPE histograms at (a) 3V, (b) 4V, and (c) 5V OV before (blue) and after (red) the 3-months run period and their Gaussian fit envelope. Inset of each plot depicts the mean numbers of photoelectrons using Gaussian envelope or Poisson statistics, $N_{pe}$(envelope) or $N_{pe}$(Poisson), respectively.

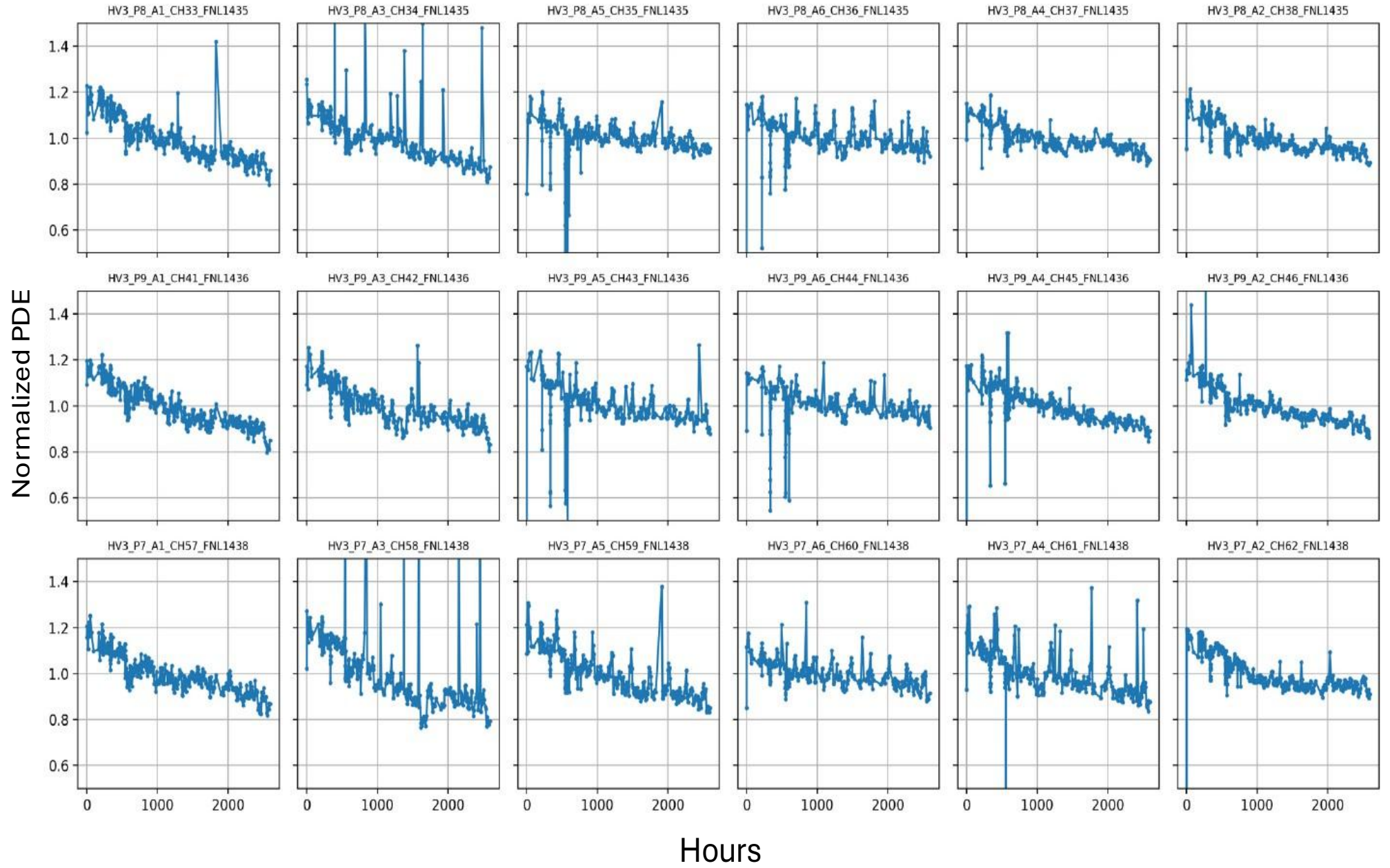

**Figure 17.** A selected set of normalized PDE using Poisson statistics for all 18 SiPMs biased at 5V OV over the 110 days test period. Some outliers occurred due to unfavorable Gaussian fits of the 0-p.e. peak on certain channels.

## 4. Room Temperature Evaluation

For large-scale photon detectors operating in cryogenic noble liquids, it is imperative to ensure the functional integrity of the entire detection system prior to filling with purified noble liquids; once submerged, the system becomes nearly non-serviceable. Monitoring only the dark rate of each photon channel confirms electronics connectivity but does not verify the single-photon response, which is the primary objective of the SiPMs.

At room temperature, however, the SiPM dark current at nominal OV is typically ~μA, see Fig. 3(b), which can lead to intolerable signal pile-up on the ASIC. Although our LArASIC design can handle relatively high leakage currents, it cannot support all 54 SiPM channels simultaneously at room temperature. Here, we demonstrate the room-temperature readout of single photons from one group of 18 SiPMs using our weak capacitive coupling concept. We utilized an optimized 1 MΩ series resistor for each SiPM; this limits the propagation of high-voltage ripple noise and accommodates the voltage drop induced by the ~μA dark current. All room temperature runs were performed following the completion of the three-month long $LN_2$ testing.

Figure 18(a) depicts a representative set of photoelectron histograms measured at room temperature at OV of 0.35 V, 0.68 V, 0.94 V, and 1.16 V. Solid dots represent the data, dotted lines are connected for visual clarity, and solid lines indicate the suggested SPE peaks. We note that the LED light pulse intensity was ~2 times higher than in the corresponding $LN_2$ runs. Single photons were successfully measured simultaneously across all 18 devices. The devices responded linearly to different photon intensities, as shown in Fig. 18(b), where the 'low intensity' corresponds to the illumination level used in the $LN_2$ runs. The statistical mean number of detected photoelectrons (relative PDE) as a function of OV is shown in Fig. 18(c). A linear dependence was observed over this small OV range, confirming the operational health of the photon readout at room temperature.

By managing data throughput and avoiding the saturation of charge-sensitive preamplifier embedded in the LArASIC, single-photon readout can be performed sequentially for each detector section. Thus, we provide a diagnosis tool for both the electronics and the photon response of large-scale photon detectors before they reach the point of irreversible deployment.

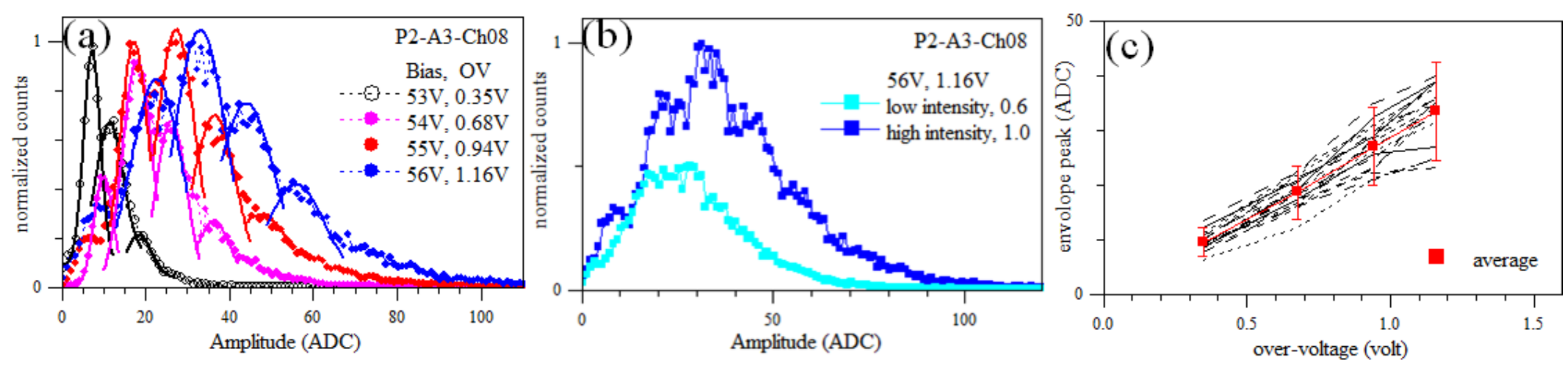

**Figure 18**. A representative set of photoelectron histograms measured at room temperature for (a) various bias OV, (b) with different LED illumination intensities, and (c) relative PDE vs OV for all 18 devices.

## 5. Conclusions

We successfully monitored selected SiPM parameters in $LN_2$ continuously for over 110 days, generating ~9 TB of waveform data. While a few SiPMs exhibited atypical responses, they remained stable throughout the study. Generally, SiPM parameters increased nominally with OV, except for the 1-p.e. energy resolution and the SNR, see Table 1. Although some channels showed ~1σ variations over the three-month test period, the DCR appeared to drop continuously by ~1σ across all SiPMs after 110 days, accompanied by a slight, gradual increase in correlated noise (CN). Notably, there was no evidence of drift being enhanced when operating at 4 V or 5 V OV compared to 3 V.

No charge gain variation was observed in the dark count data. However, we initially observed a charge gain oscillation with a periodicity of ~11 days in the LED data, which was attributed to a long-term phase drift of the 10 MHz reference clock. After applying a robust waveform-fitting algorithm to all LED data, we concluded that no physical charge gain drift occurred; charge gain remained stable at all bias levels over the three-month period. Conversely, a discernible PDE drop of ~5% per month was observed across all channels, which warrants further investigation. Table 1 summarizes these findings.

Table 1: SiPM results normalized to 3V over-voltage and average of all $LN_2$ runs

| Section # | Over-voltage | 3V | 4V | 5V | 3-month drift |
|---|---|---|---|---|---|
| 3.1 | DCR | 1 | 1.19 [n/c] | 1.39 [n/c] | - 1 σ |
| 3.2 | 1-pe resolution | 1 | 0.93 [n/c] | 0.87 [n/c] | - 1 σ |
| 3.3 | SNR | 1 | 1.27 [n/c] | 1.22 [n/c] | < 1 σ |
| 3.4 | CN (pe) | 1 | 1.28 [n/c] | 1.66 [n/c] | +1 σ |
| | CT probability | 1 | 1.35 [1.32] | 1.50 [1.65] | n/c |
| 3.5 | Charge gain | 1 | 1.27 [1.33] | 1.57 [1.66] | < 1 σ |
| 3.6 | PDE | 1 | 1.27 [1.43] | 1.22 [1.85] | few % drop/month |

*[HPK spec. at room temp.]*

This photon readout approach, utilizing the weak-coupling concept with the LArASIC, can be implemented for SiPM with significantly larger capacitance. To optimize the dynamic range of the signal charge, we utilized the lowest ASIC gain (4.3 mV/fC), the smallest decoupling capacitor (10 pF), and a 1 μs peaking time. All of these parameters can be dynamically adjusted to optimize the performance for various SiPM arrays with different characteristics. Finally, we demonstrated that our system can simultaneously read out single photons from a selected group of SiPMs at room temperature with photon-number-resolving capability.

**Acknowledgments**

This work was supported in part by U.S. Department of Energy under Contract No. DE-AC02-98CH10886. We gratefully acknowledge the cooperation of the DUNE collaboration in providing the SiPMs for testing. We thank our coordinator Dr. Peter Shanahan and this comments. We appreciated the technical assistance of Jack Fried, Donald Pinelli, Joseph Pinz, Antonio Verderosa, Wenrui Ray Lin, and Bill Smith.